\documentclass[journal=jpclcd,manuscript=letter]{achemso}
\usepackage{chemformula} 
\usepackage[T1]{fontenc} 
\usepackage{soul}

\graphicspath{{figs_v0.4/}}

\newif\ifnohighlight
\nohighlightfalse
\renewcommand\hl[1]{%
	\bgroup
	\hskip0pt\color{green!50!black}%
	#1%
	\egroup
}
\nohighlighttrue 
\ifnohighlight
\renewcommand\hl[1]{#1}
\fi

\newcommand{\CEISAM}{Nantes Universit\'e, CNRS,  CEISAM UMR 6230, F-44000 Nantes, France}
\newcommand{\IUF}{Institut Universitaire de France (IUF), F-75005 Paris, France}

\author{Thomas V. Papineau}
    \affiliation[UN, Nantes]{\CEISAM}
\author{Denis Jacquemin}
    \affiliation[UN, Nantes]{\CEISAM}
    \alsoaffiliation[IUF]{\IUF}
\author{Morgane Vacher}
    \affiliation[UN, Nantes]{\CEISAM}
    \email{Morgane.Vacher@univ-nantes.fr}

\title[An \textsf{achemso} demo]
  {Which Electronic Structure Method to Choose in Trajectory Surface Hopping Dynamics Simulations? Azomethane as a Case Study}

\begin{document}

\begin{tocentry}




    \includegraphics[width=1.0\textwidth]{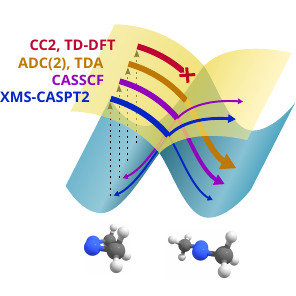}

\end{tocentry}

\begin{abstract}
Non-adiabatic dynamics simulations have become a standard approach to explore photochemical reactions.
Such simulations require underlying potential energy surfaces and couplings between them, calculated at a chosen level of theory, yet this aspect is rarely assessed.
Here, in combination with the popular trajectory surface hopping dynamics method, we use a high-accuracy XMS-CASPT2 electronic structure level as a benchmark for assessing the performances of various post-Hartree-Fock methods (namely CIS, ADC(2), CC2 and CASSCF) and exchange-correlation functionals 
(PBE, PBE0, CAM-B3LYP) in a TD-DFT/TDA context, using the isomerization around a double bond as test case.
Different relaxation pathways are identified, and the ability of the different methods to reproduce their relative importance and timescale is discussed. The results show that multi-reference electronic structure methods should be preferred\hl{, when studying non-adiabatic decay between excited and ground states}. If not affordable, TD-DFT with TDA and hybrid functionals, and ADC(2) are efficient alternative, but overestimate the non-radiative decay yield and thus may miss deexcitation pathways.
\end{abstract}


Light-induced phenomena, such as photochemical reactions, have widespread applications
ranging from biology to energy conversion~\cite{Wald-1968,Liu-1990,Polosukhina-2012,Zhang-2013,Richards-2016,Baroncini-2020}.
The photo-induced events involved in these phenomena
generally occur on the femtosecond timescale, making them ultrafast phenomena, a challenge for experiments.
In order to obtain theoretical insights into these processes, nonadiabatic dynamics simulations are the methods of choice~\cite{Vacher-2016-TCA,Curchod-2018,Crespo-Otero-2018}.
Among them, the most popular is trajectory surface hopping (TSH)\cite{Tully-1971,tullyMolecularDynamicsElectronic1990}, an on-the-fly mixed quantum-classical method. TSH uses the quantum propagation of an electronic wavepacket, inducing ``hops'' of independent semiclassical trajectories between the electronic states. It is an \textit{on-the-fly} dynamics method, meaning that the potential energy surfaces (PESs) are generated, using a selected electronic structure method,  as needed as the nuclear trajectories move along.

During non-radiative relaxations, the molecule encounters regions with strongly non-adiabatically coupled electronic states, such as the so-called conical intersections~\cite{Teller-1937,Robb-1995,Yarkony-2001,Domcke-2012}. 
In general, the most suited electronic structure methods for describing such intersections are multi-configurational wavefunction methods~\cite{Roos-1980,Yeager-2008,Olsen-2011} -- although other methods were shown to be adequate in specific cases\cite{Tuna-2015}.
However, the multi-configurational methods come with a high computational cost, scaling exponentially with the size of the active space.
This limits the theoretical studies to small molecular systems (few tens of atoms) and/or short simulation time (up to a ps).
To overcome this bottleneck, the alternative has been to use computationally cheaper single-reference electronic structure methods such as the second-order algebraic diagrammatic construction (ADC(2))~\cite{Plasser-2014,Xie-2019,Braun-2022,Mansour-2022,Mansour-2022-JPCL,Ferte-2023}, or linear response time-dependent density functional theory (TD-DFT)~\cite{Plasser-2014,Ye-2020,Cao-2020,Papai-2021,Liu-2021,Zobel-2023}, sometimes with the Tamm-Dancoff approximation (TDA)~\cite{Talotta-2020} or its spin-flip variant~\cite{Minezawa-2019}. Previous works studied the effect of the electronic structure methods onto the crossing topologies\cite{Gozem-2014}.

A common approach in the literature to benchmark electronic structure methods is to compare energies and geometries of key structures or along specific cuts of the potential energy surfaces\cite{Tuna-2015,Marsili-2021}. But globally, the impact of the choice of the electronic structure methods on the non-adiabatic dynamics has been little studied\cite{Barbatti-2012,Mai-2019,Chakraborty-2021,Janos-2023,Barneschi-2023}, since simulating and comparing dynamics with several electronic structure approaches is often computationally beyond reach. \hl{Very recently, the nonadiabatic dynamics of cyclopropanone was extensively studied with different electronic structure methods and nonadiabatic dynamics algorithms, showing a much larger sensitivity to the former than the latter\cite{Janos-2023}.}
The goal of the present Letter is to benchmark the suitability for non-adiabatic dynamics simulations, in particular internal conversion, of a series of different electronic structure methods: (extended) multi-state complete active space second order perturbation theory (MS-CASTP2~\cite{Andersson-1990,Andersson-1992,Finley-1998} and XMS-CASPT2~\cite{Shiozaki-2011}), state-averaged complete active space self-consistent field (SA-CASSCF)~\cite{Olsen-2011,Roos-1980}, second-order approximate coupled cluster singles and doubles (CC2)~\cite{Christiansen-1995}, ADC(2)~\cite{Schirmer-1982,Schirmer-2004}, configuration interaction singles (CIS)~\cite{Bene-1971,Foresman-1992}, TD-DFT~\cite{Runge-1984,Casida-1995} with several functionals -- PBE,\cite{Per96} PBE0\cite{Ada99} and CAM-B3LYP,\cite{Yan04} as respective representatives of local, global hybrid and range-separated hybrid models -- with and without TDA~\cite{Hirata-1999}. The formal scaling of the CC2, ADC(2) and CIS methods is $N^5$, with $N$ the number of atoms, while the TD-DFT approaches formally scale to $N^4$.
Our aim is to pinpoint a light electronic structure scheme allowing the study larger molecules of chemical interest (several hundreds of atoms) and/or on a longer scale (tens to hundreds of ps). 

To assess the performances of a large panel of electronic structure methods, this Letter focuses on the \textit{cis}-to-\textit{trans} photoisomerization of the azomethane molecule (Scheme~\ref{sch:AZM}). 
This photochemical reaction is known to occur rapidly in approximately 200~fs upon light excitation to its first electronic state ($S_1$, $n\pi^*$) \cite{sellnerPhotodynamicsAzomethaneNonadiabatic2010,Merritt-2023}. 
It proceeds through a rapid rotation about the central CNNC moiety, reaching a $S_1$/$S_0$ conical intersection located around CNNC = 90$^{\circ}$, roughly halfway between the \textit{cis} and \textit{trans} isomers, at approximately $t$ = 40 fs.
Previous simulations have made the distinction between ``standard'' trajectories, which decay through this first conical intersection, and ``rotator'' trajectories which decay through a second symmetry equivalent intersection reached after a 270$^{\circ}$ total rotation.\cite{sellnerPhotodynamicsAzomethaneNonadiabatic2010,Merritt-2023}
 \hl{The effect of the initial conditions on the longer-timescale dynamics was recently investigated in that system\cite{Pieroni-2023}.}
\textit{Cis}-\textit{trans} isomerization being one of the main types of photochemical reactions, this photoisomerization reaction provides an ideal playground for studying the behavior of different electronic structure methods in describing excited state population decay and geometrical changes.
The small size of the chosen system allows us to use a high-accuracy reference method accounting for both static and dynamic electronic correlations, namely XMS-CASPT2, and to propagate the electronic wavefunction with explicitly calculated non-adiabatic coupling vectors (NACV) (see Methods section for details).

\begin{scheme}
	\includegraphics[width=.5\textwidth]{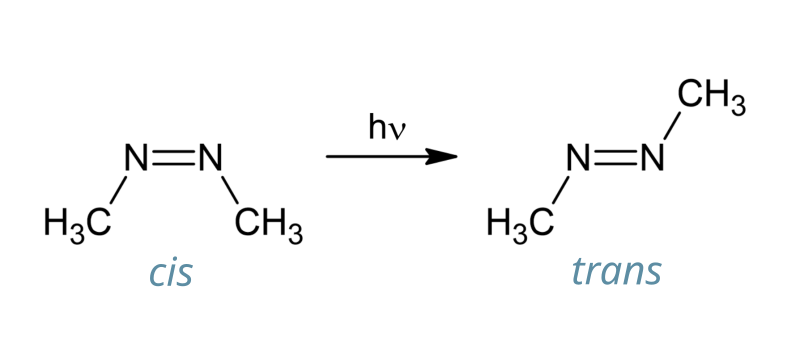}
	\caption{\textit{cis}-to-\textit{trans} photoisomerization of azomethane upon excitation to its first electronic excited state.}
	\label{sch:AZM}
\end{scheme}
 
\begin{figure}
    \includegraphics[width=.5\textwidth]{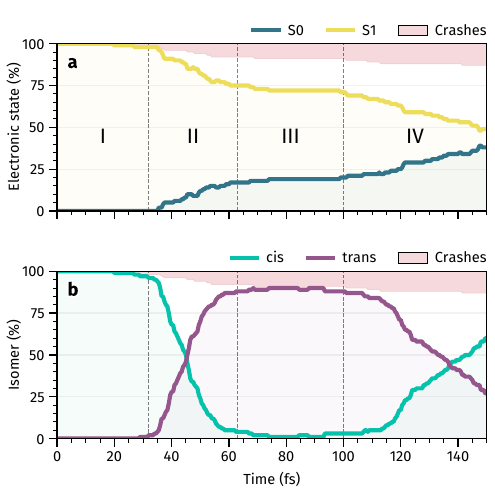}
    \caption{Time evolution of (a) electronic populations of $S_0$ (blue) and $S_1$ (yellow) states, and (b) \textit{cis} (teal) and \textit{trans} (purple) isomer populations during the XMS-CASPT2 (NACV) reference dynamics. Crashed trajectories are represented with the red area. The four different time regimes, namely I, II, III and IV, are delimited by dashed lines.}
    \label{fgr:ref_pops}
\end{figure}

Figures~\ref{fgr:ref_pops}a and~\ref{fgr:ref_pops}b show the electronic and nuclear dynamics, respectively, obtained with the reference electronic structure method. 
Nuclear geometries are assigned to \textit{trans} or \textit{cis} isomers depending on their CNNC dihedral angle: $>$ or $<$ 90$^{\circ}$, respectively.
One can distinguish four time regimes across the simulation: 
	I) an initial delay of about 30~fs during which all molecules remain \textit{cis} and in the first excited state; 
	II) the crossing of a first conical intersection region up to approximately 60~fs, inducing a non-adiabatic decay of about 20-25\% from $S_1$ to $S_0$, associated with a complete isomerization from \textit{cis} to \textit{trans}; 
	III) a second plateau in the electronic populations lasting approximately 40 fs, while most molecules evolve in the \textit{trans} configuration; and 
	IV) the crossing of a second strong-coupling region with further non-adiabatic decay until the end of the simulation, during which a fraction of the molecules come back to the \textit{cis} configuration.

\begin{figure}
	\includegraphics[width=1.0\textwidth]{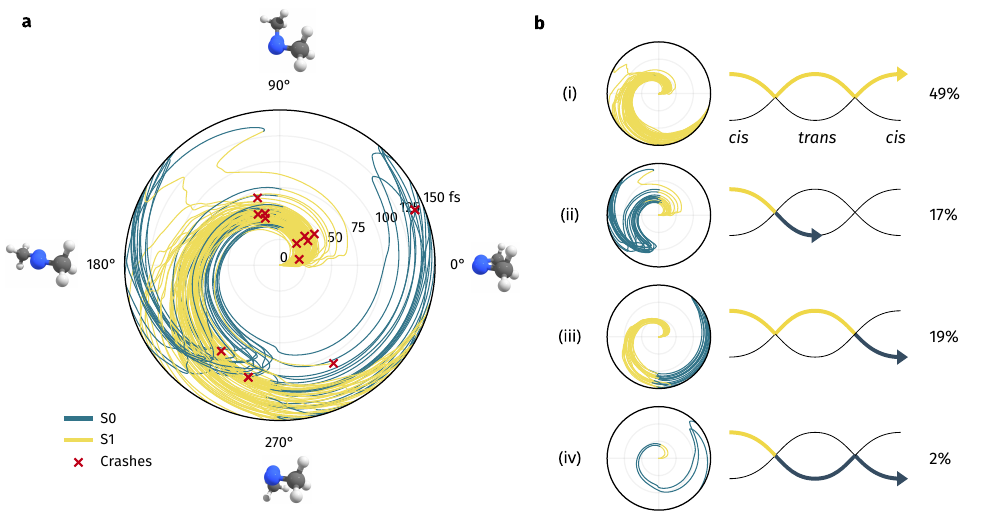}
    \caption{(a) Polar representation of all 100 trajectories of the XMS-CASPT2 (NACV) reference dynamics. For each trajectory, the evolution of the CNNC dihedral angle is represented by the rotation around the plot, while the elapsed simulation time is depicted by the increasing distance from the center. The active state is indicated by the blue ($S_0$) or yellow ($S_1$) color. Crashes of trajectories are marked with a red cross. (b) Separation into the four main deexcitation pathways. }
    \label{fgr:ref_polar}
\end{figure}

Figure~\ref{fgr:ref_polar}a shows, with a polar graph, the time evolution of the CNNC dihedral angle -- the representative coordinate of the isomerization --  with the time indicated by the increasing distance from the center, for all 100 trajectories. 
The active state on which each nuclear trajectory is propagated is indicated by the color of the curve.
Analyzing both electronic and geometrical information in such a way allows identifying the four main deexcitation pathways presented in Figure~\ref{fgr:ref_polar}b: 
	(i) trajectories staying on $S_1$ until the end of the simulation time; 
	(ii) trajectories deexciting toward $S_0$ at the first conical intersection (dihedral of 90$^{\circ}$), ending in \textit{trans}; 
	(iii) trajectories staying excited on $S_1$ at the first conical intersection, then decaying toward $S_0$ at the second intersection occurrence (270$^{\circ}$), thus ending in \textit{cis}; 
	(iv) trajectories deexciting at the first conical intersection and having enough kinetic energy to cross the isomerization barrier in the ground state, thus ending in \textit{cis}. 
In the language introduced above, pathways (ii) and (iv) describe ``standard'' trajectories, while pathway (iii) describes ``rotator'' trajectories. 
In the reference XMS-CASPT2 dynamics, pathway (i) dominates with almost half of the trajectories; pathways (ii) and (iii) represent each a bit less than 20\% of the trajectories; pathway (iv) is trifling.

Let us now turn our attention to the dynamics obtained with the other methods. 
First, staying with CASPT2, two other electronic propagation algorithms are compared: the Local Diabatization (LD),\cite{Granucci-2001} and a simpler energy-gap threshold-based approach (thrs), in which hops toward $S_0$ are enforced when the $S_1-S_0$ energy gap goes below 0.1 eV. 
XMS-CASPT2 with LD was performed using the BAGEL code\cite{BAGEL}. 
XMS-CASPT2 and MS-CASPT2 with LD were also achieved using OpenMolcas\cite{OpenMolcas-2019,OpenMolcas-2023}: there, the electronic wavefunction overlap required in the LD algorithm is calculated approximately, taking into account the CASSCF part of the CASPT2 wavefunctions only, i.e., neglecting the perturbation part. 
It is emphasized that this approximate LD scheme, noted ``lLD'' hereafter, uses the correct CASPT2 state ordering and can describe (if necessary, based on the CASPT2 results) the electronic states as linear combinations of CASSCF wavefunctions.
Despite the development of NACV for methods like TD-DFT\cite{Send-2010,Tavernelli-2010,Li-2014,Ou-2015}, the threshold-based procedure remains the most commonly used for TD-DFT and ADC(2) (to describe  $S_1\rightarrow S_0$ transitions).
The other benchmarked electronic structure methods are CASSCF, CC2 and ADC(2), CIS, TD-PBE, TD-PBE0, TD-CAM-B3LYP, TDA-PBE, TDA-PBE0 and TDA-CAM-B3LYP. 
Additional computational details about each method are provided in the Methods section.
Importantly, some electronic structure methods are more prone to failure, leading to crashes, than others.
This typically appears due to convergence errors in the vicinity of electronic degeneracies. In the present work, all crashed trajectories are kept and considered as a special case. 
 
\begin{figure}[ht]
    \includegraphics[width=.5\textwidth]{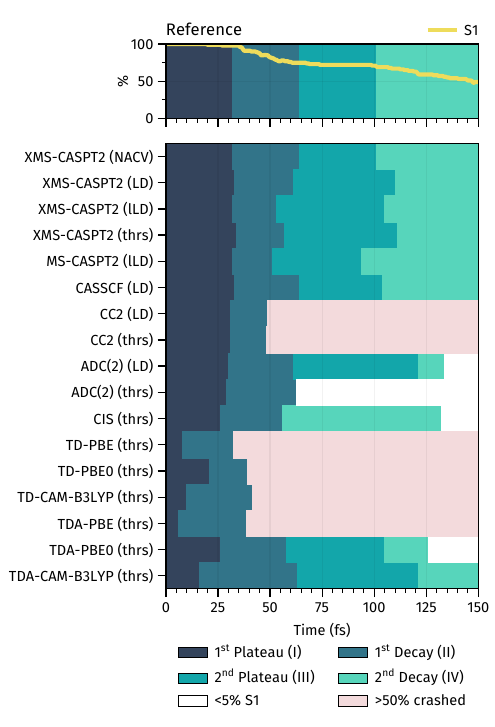}
    \caption{Durations of the four regimes identified during the non-adiabatic dynamics calculated with the reference XMS-CASPT2 (NACV) method (reminded at the top). 
    White areas indicate time windows where less than 5\% of the initial $S_1$ population is remaining (because of deexcitations and/or crashes), making the regime no longer identifiable. 
    Red areas indicate a regime with more than 50\% of the trajectories having crashed.}
    \label{fgr:comparison_timescales}
\end{figure}

Figure~\ref{fgr:comparison_timescales} compares the time duration of the four different regimes identified in the reference XMS-CASPT2 (NACV) dynamics (Figure~\ref{fgr:ref_pops}) for all benchmarked methods.
All multi-configurational methods, including CASSCF, accurately reproduce both the four regimes observed in the reference and their duration.
The agreement between NACV and LD schemes was already demonstrated in an extensive benchmark studies of the different electronic propagation schemes\cite{Merritt-2023}. 
The results of lLD indicate that considering only the CASSCF part of the CASPT2 wavefunction for calculating overlaps is a reasonable approximation in the present case, despite a slight underestimation of the duration of the first decay (regime II).
Importantly, the threshold-based XMS-CASPT2 dynamic shows little difference when compared to the XMS-CASPT2 (NACV) dynamics, justifying the use of this procedure for CIS and TD-DFT methods.

Regarding the single-reference methods, CC2 appears to be highly unstable near the $S_1/S_0$ intersections in azomethane, resulting in systematic failures in their vicinity. This result is in agreement with a previous study devoted to the non-adiabatic dynamics among quasi-degenerate excited states in adenine.\cite{Plasser-2014} In contrast, CC2 was able to predict expected characteristics of the $S_1/S_0$ conical intersection in a model of rhodopsin protein chromophore\cite{Tuna-2015}. In the present work, all tested functionals without applying TDA, as well as TDA-PBE, show a similarly unstable behavior. 
This results in prematurely ended dynamics after regime II for these methods. The use of TDA proved more resilient and yielded no significant failure. 
This observation is in agreement with a previous study combining TSH and TD-DFT.\cite{Tapavicza-2008}
ADC(2) reproduces well the four time regimes and their duration. CIS predicts a continuous decay in the $S_1$ population until the end of the simulation, although with a lower slope during regime IV. 
For all tested functionals (applying or not TDA), the initial delay preceding the first decay (regime I) is significantly too short -- PBE0 performs slightly better in that regard.
This is compensated by a longer regime II, making the end of this regime appearing at ca.~60 fs irrespective of the chosen electronic structure method (when there is not too many crashes).
Both TDA-PBE0 and TDA-CAM-B3LYP reproduces well the four different time regimes; the timescales obtained with TDA-PBE0 are in better agreement with those of the reference.
While CASSCF is the best performing method in reproducing reference's timescales, ADC(2) is notably efficient. 
Moreover, TDA along with an hybrid functional provides acceptable results despite the too short initial delay.

\begin{figure}
	\includegraphics[scale=1]{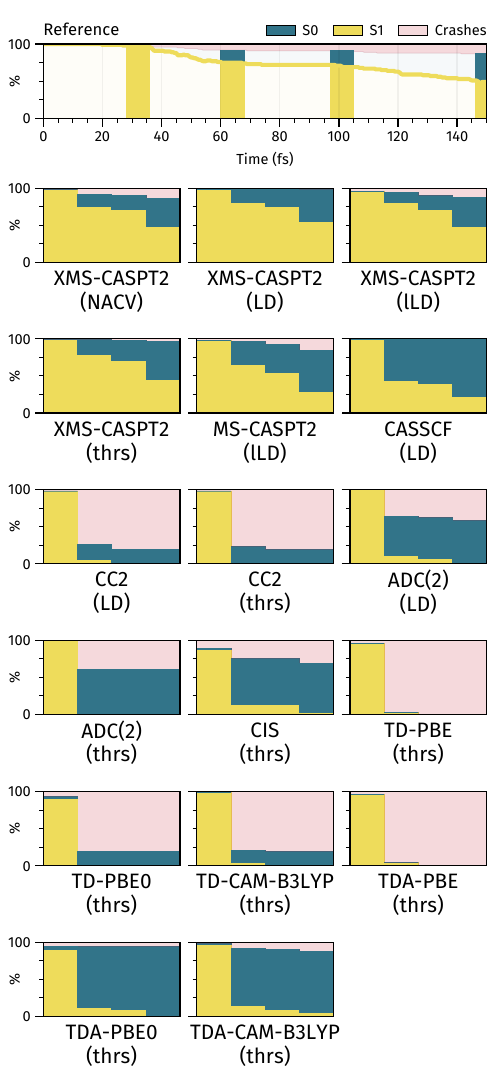}
	\caption{Electronic state populations ($S_0$ in blue, $S_1$ in yellow) at the end of each identified time regime, obtained with all electronic structure methods. Crashed trajectories are represented by a red area.}
    \label{fgr:comparison_populations}
\end{figure}

Figure~\ref{fgr:comparison_populations} compares the yield of the non-adiabatic decay by extracting the electronic population of both states at the limits of the regimes I-IV identified with the reference XMS-CASPT2 (NACV) results.  
The exact times limiting the different regimes depend on the electronic structure method (see Figure~\ref{fgr:comparison_timescales}). 
In particular, CIS and ADC(2) (thrs) have no regime III.
The decay yield is well reproduced by XMS-CASPT2 with LD and lLD electronic propagation schemes. 
It is slightly faster in the case of the threshold-approach. 
This can be explained: the threshold-based hopping procedure does not allow trajectories to hop back to the excited state (back-hopping phenomenon).
As seen with the timescale analysis, most of the trajectories crashed with CC2, PBE (with and without TDA) and with the hybrid functionals without TDA. 
All other tested electronic structure methods present larger decays than the reference one. 
While the use of the threshold-approach can explain part of the too important decay, it cannot account for all of it.
The too fast decay predicted here in azomethane by TDA-PBE0 and TDA-CAM-B3LYP contrasts with the failure of TD-DFT to describe the non-adiabatic decay in adenine studied in previous works.\cite{Barbatti-2012,Plasser-2014}
Ignoring MS-CASPT2, CASSCF is here again the best performing method in comparison to XMS-CASPT2. 
While showing different amounts of failure, the single-reference methods ADC(2) (LD), CIS, TDA-PBE0 and TDA-CAM-B3LYP predict similarly too large non-adiabatic decay yields. 

\begin{figure}[htp]
	\includegraphics[scale=1]{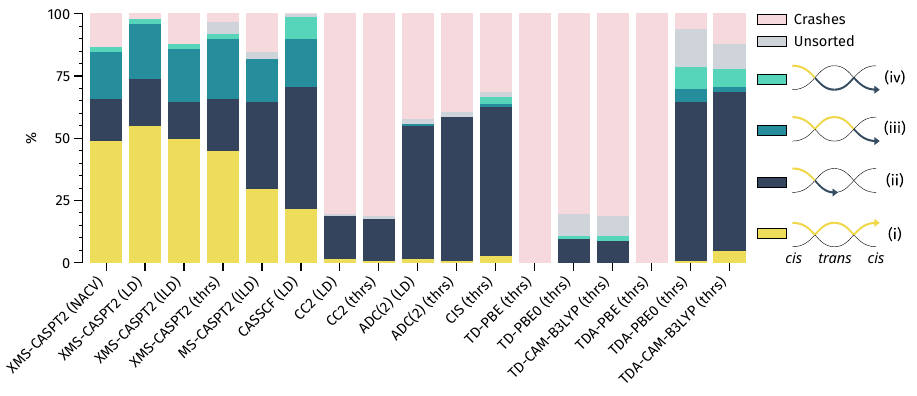}
	\caption{Relative importance of the four identified relaxation pathways (i-iv) obtained with the different electronic structure methods.}
    \label{fgr:comparison_pathways}
\end{figure}

Figure~\ref{fgr:comparison_pathways} presents the relative importance of the main relaxation pathways identified in the reference dynamics (Figure~\ref{fgr:ref_polar}) for each studied electronic structure method. 
It is noted that for some methods, mainly DFT-based ones, a significant amount of trajectories do not follow any of the four typical relaxations pathways and thus remain unsorted. 
All multi-reference wavefunctions methods reproduce qualitatively the main relaxation pathways, with MS-CASPT2 and CASSCF overestimating the importance of the ``standard'' trajectories, that is, pathways (ii) and (iv). 
This is in agreement with the too efficient electronic population decay through the first encountered conical intersection (Figure~\ref{fgr:comparison_populations}). 
This feature is even more present in the single-reference post-HF wavefunction and DFT methods: with them, the pathways that stay on the excited state after the first crossing are incorrectly negligible and pathway (ii) dominates largely the dynamics. 
One could be tempted to assign this result to the threshold approach. 
However, the XMS-CASPT2 (thrs) results indicate that it is not the main reason and that this feature is rather an intrinsic failure of the single-reference approaches.

In this Letter, we have tested a range of different electronic structure methods for simulating the \textit{cis}-to-\textit{trans} photoisomerisation reaction of azomethane in combination with the surface hopping dynamics method: multi-reference (XMS-CASPT2, MS-CASPT2 and CASSCF) and single-reference (ADC(2), CC2, CIS) wavefunction methods as well as TD-DFT based approaches (PBE, PBE0 and CAM-B3LYP functionals with and without TDA). 
Different electronic propagation and hopping algorithms were  considered: using NACV, LD or the energy-threshold approach. 
During the first 150~fs after photoexcitation to the $S_1$ state, the molecule has time to reach the first 90$^{\circ}$ conical intersection after ca.~40~fs, and the 270$^{\circ}$ conical intersection from 120~fs and onwards. 
Analyzing the electronic and geometric time evolutions, four different time regimes and relaxation pathways could be identified in the reference XMS-CASPT2 (NACV) dynamics.

Multi-reference methods reproduce closely the timescale and yield of the induced non-adiabatic dynamics, as well as the different relaxation pathways.
CASSCF slightly overestimates the decay through the first conical intersection, and thus the amount of ``standard'' trajectories and corresponding pathways. 
CC2, TD-PBE, TDA-PBE, TD-PBE0 and TD-CAM-B3LYP appear to be too unstable in the vicinity of electronic degeneracies and thus, no trustworthy statistical results could be obtained. 
ADC(2) (LD), TDA-PBE0 and TDA-CAM-B3LYP  qualitatively reproduce the timescale of the decay (despite crashes of up to 40\% of the trajectories) but predicts an almost total decay through the first crossing and thus poorly describe  the ``rotator'' trajectories. 
CIS, in addition to underestimating the ``rotator'' trajectories, does not reproduce the plateau in the electronic population evolution after crossing the first  intersection. 
The too important non-adiabatic decay obtained with these methods cannot be attributed to the threshold approach.

In short, the present trajectory surface hopping work on azomethane shows that multi-reference electronic structure methods are the preferred choices to simulate non-adiabatic dynamics. If not affordable, ADC(2), TDA-PBE0 and TDA-CAM-B3LYP are alternatives which reproduce well the timescale of the reference dynamics but overestimate the non-adiabatic decay yield when going through a crossing and thus may miss some deexcitation pathways. Interestingly, the suitability of the different electronic structure methods could hardly be predicted from a static analysis of the potential energy surfaces (see SI for energetics at Franck-Condon geometry and scan along the CNNC dihedral angle coordinate). This suggests that the common approach to benchmark and validate electronic structure methods before simulating dynamics may not be sufficient\hl{, as concluded in a recent work\cite{Janos-2023}}. Of course, the present conclusions were obtained in the specific case of the \textit{cis}-to-\textit{trans} photoisomerisation reaction of azomethane upon $S_1$ excitation, and thus cannot be assumed to be general for all molecules, all electronic states, nor all types of photo-induced processes.

\section*{Methods}

The surface hopping simulations were carried out with the SHARC code\cite{SHARC3.0,Mai-2018-SHARC} using the fewest switches algorithm: 100 trajectories were simulated for 150 fs with a time step of 0.5 fs. 
Initial conditions for the different sets of 100 trajectories were generated with SHARC tools, sampling geometries and velocities from the Wigner distribution (without temperature broadening). This was done by using harmonic frequencies calculated at the relevant optimized ground-state geometry using the respective electronic structure level. The initial conditions are thus in principle not identical for the various tested electronic structure methods, although in practice no significant difference was observed (see SI). This choice was made in order to reflect a realistic non-adiabatic study. \hl{We, nevertheless, simulated dynamics at ADC(2) and TDA-PBE0 levels of theory with initial conditions generated with XMS-CASPT2 method (see SI): the results show that the initial conditions have little effect (in contradiction with a previous study on that system\cite{Pieroni-2023}) and confirm that the difference come from the electronic structure.}
To conserve the total energy after a hop, the full velocity vector was rescaled. An energy based decoherence correction was applied with a decay factor of 0.1 Hartree\cite{Granucci-2007}. For all electronic structure methods benchmarked herein, the selected basis set is \emph{def2}-SVP\cite{Weigend-2005}.

The reference XMS-CASPT2 (NACV) calculations have been achieved with the BAGEL code\cite{BAGEL} and a modified version of the SHARC-BAGEL interface, using an active space consisting of 6 electrons in 4 orbitals: the $\pi$, $\pi^*$ and the two N lone pairs. 
No shift was used, and IPEA was deactivated. BAGEL was also used to perform the XMS-CASPT2 (LD) simulations.
XMS-CASPT2 (lLD), MS-CASPT2 and CASSCF simulations were performed with OpenMolcas\cite{OpenMolcas-2019,OpenMolcas-2023}, using the same active space and RI approximation with analytical Cholesky decomposition.
CC2 and ADC(2) electronic structure calculations have been done with the Turbomole 7.3 code\cite{TURBOMOLE}, using the defaults parameters requested by the SHARC-RICC2 interface (RI approximation with default selection of auxilliary basis set, and frozen core orbitals).
Finally, CIS, TD-DFT and TDA-DFT calculations were made with Gaussian 16 rev.~A03,\cite{g16} using the \textit{tight} SCF convergence criteria and \textit{fine} integration grid. CIS calculations have been achieved by requesting TDA-HF calculations.
The TD- and TDA-PBE0 dynamics have also been performed using Turbomole 7.3 code\cite{TURBOMOLE} (using a modified version of the SHARC-RICC2 interface) and Orca 5.0.3\cite{ORCA,ORCA5}. In Turbomole, the RI approximation was activated and the \textit{m3} integration grid was used. Within Orca, the RIJCOSX approximation was enabled, along with the \emph{def2}/J and \emph{def2}-SVP/C auxilliary basis sets. The \textit{tight} SCF convergence criteria was requested and the \textit{defgrid2} integration grid was selected.

\begin{acknowledgement}
The authors thank the ANR for financial support in the framework of the BiBiFlu project.
The project is also partly funded by the European Union through ERC grant 101040356 (M.V.). Views and opinions expressed are however those of the author(s) only and do not necessarily reflect those of the European Union or the European Research Council Executive Agency. Neither the European Union nor the granting authority can be held responsible for them. The authors thank the CCIPL/Glicid mesocenter installed in Nantes and GENCI-IDRIS (Grant 2021-101353) for generous allocation of computational time.
\end{acknowledgement}

\begin{suppinfo}
Active space orbitals. Comparison of initial internal coordinates for all electronic structure methods. \hl{Test of the initial conditions on the dynamics.} Total energy variations along the dynamics for all benchmarked electronic structure methods. Time evolutions of electronic populations and isomer populations defining the different time regimes for all benchmarked electronic structure methods.  Energetics at Franck-Condon geometry and scan along the CNNC dihedral angle coordinate, with all tested electronic structure methods. The modified SHARC interfaces can be made available upon reasonable request.


\end{suppinfo}

\bibliography{bibliography}

\end{document}